\documentclass[letterpaper, 10 pt, conference]{ieeeconf}
\IEEEoverridecommandlockouts
\usepackage{amsmath,latexsym,amssymb,multicol}
\usepackage[pdftex]{graphicx}
\usepackage{blkarray}
\newcommand{\diag}{\ensuremath{\mathrm{diag}}}

\newcommand{\Ima}{\ensuremath{\mathrm{Im}}}
\newcommand{\Ker}{\ensuremath{\mathrm{Ker}}}
\newcommand{\Span}{\ensuremath{\mathrm{span}}}
\newtheorem{theorem}{Theorem}
\newtheorem{lemma}{Lemma}
\newtheorem{coro}{Corollary}
\title{\Large{\textbf{The Kalman Decomposition for Linear Quantum Stochastic Systems}}}

\author{Symeon Grivopoulos$^1$ \and Guofeng Zhang$^2$ \and Ian R. Petersen$^1$ \and John Gough$^3$%
\thanks{$^1$ Symeon Grivopoulos and Ian R. Petersen are with the School of Engineering and Information Technology, UNSW Canberra, Canberra BC 2610, Australia {\tt\small symeon.grivopoulos@gmail.com, i.r.petersen@gmail.com}}
\thanks{$^2$ Guofeng Zhang is with the Department of Applied Mathematics, The Hong Kong Polytechnic University, Hong Kong {\tt\small Guofeng.Zhang@polyu.edu.hk}}
\thanks{$^3$ John Gough is with the Department of Physics, Aberystwyth University, Wales,cSY23 2BZ, Aberystwyth, UK {\tt\small jug@aber.ac.uk}}
\thanks{This work was supported by the Australian Research Council under grant FL110100020}}

\begin{document}

\maketitle

\begin{abstract}
The Kalman decomposition for Linear Quantum Stochastic Systems in the real quadrature operator representation, that was derived indirectly in \cite{zhagripet16} by the authors, is derived here directly, using the ``one-sided symplectic'' SVD-like factorization of \cite{xu03} on the observability matrix of the system.
\end{abstract}

\section{Introduction}
\label{Introduction}

Linear Quantum Stochastic Systems (LQSSs) are a class of models used in linear quantum optics \cite{garzol00,walmil08,wismil10}, circuit QED systems \cite{matjirper11, kerandku13}, quantum opto-mechanical systems \cite{tsacav10, masheipir11, hammab12, donfiokuz12}, and elsewhere. The mathematical framework for these models is provided by the theory of quantum Wiener processes, and the associated Quantum Stochastic Differential Equations \cite{par99,mey95,hudpar84}. Potential applications of LQSSs include quantum information processing, and quantum measurement and control. In particular, an important application of LQSSs is as coherent quantum feedback controllers for other quantum systems, i.e. controllers that do not perform any measurement on the controlled quantum system, and thus, have the potential to outperform classical controllers, see e.g. \cite{yankim03a,yankim03b,jamnurpet08,nurjampet09,maapet11b,zhajam12,mab08,hammab12,critezsoh13}.

Controllability (stabilizability) and observability (detectability) of a classical linear system are necessary and sufficient conditions for the existence of a stabilizing controller for it, and thus, prerequisites for various control design methods. These notions, and the related mathematical concepts and techniques, can be transferred essentially unchanged to LQSSs, where, again, they are prerequisite for various design methods, see e.g. \cite{jamnurpet08,nurjampet09,pet10}. There is, however, an important difference from the classical case: The allowed state transformations in LQSSs (for the purpose of related state-space decompositions) cannot be arbitrary, but are fundamentally restricted by the laws of quantum mechanics. More specifically, in the so called \emph{real quadrature operator representation} of an LQSS that is used in this work, the only transformations that preserve its structure (see Subsection \ref{Linear Quantum Stochastic Systems}) are real symplectic ones. Recently, various investigations of controllability and observability for LQSSs have appeared in the literature, see e.g. \cite{gouzha15, gutyam16,zhagripet16}. In \cite{zhagripet16}, the authors of the present work showed that, a Kalman decomposition of a LQSS is always possible with a real orthogonal and symplectic transformation. Moreover, they uncovered the following interesting structure in the decomposition: The controllable/observable ($co$), and uncontrollable/unobservable subsystems ($\bar{c}\bar{o}$) are LQSSs in their own right, as is to be expected from a physics perspective. Furthermore, the states of the controllable/unobservable ($c\bar{o}$) subsystem are conjugate variables of the states of the uncontrollable/observable ($\bar{c}o$) subsystem. An immediate consequence of this is that, a $c\bar{o}$ subsystem exists if and only if a $\bar{c}o$ subsystem does, and they always have the same dimension. This is a consequence of the special structure of LQSSs.

The construction of the Kalman decomposition in \cite{zhagripet16}, is performed first in the so called \emph{creation-annihilation operator representation} of a LQSS, where special bases for the $co$, $\bar{c}\bar{o}$, $c\bar{o}$, and $\bar{c}o$ subspaces are constructed, and the result is then translated in the real quadrature representation. We should point out that the Kalman decomposition of a LQSS in the real quadrature representation offers an advantage over the corresponding decomposition in the creation-annihilation representation of the LQSS: In the former, the $c\bar{o}$ and $\bar{c}o$ subsystems are separate, as usual, while in the latter, the two subsystems are merged, due to the grouping of states imposed by that representation. In this work, we present a derivation of the Kalman decomposition of a LQSS, directly in the real quadrature operator representation. This derivation uses the ``one-sided symplectic'' SVD-like factorization of \cite{xu03} on the observability matrix of the LQSS, and leads directly to the desired decomposition.  Its value lies in its brevity and directness in uncovering the structure of the Kalman decomposition of LQSSs.

\section{Background Material}
\label{Background Material}

\subsection{Notation and terminology}
\label{Notation and terminology}

\begin{enumerate}
  \item $x^{*}$ denotes the complex conjugate of a complex number $x$ or the adjoint of an operator $x$, respectively. For a matrix $X=[x_{ij}]$ with number or operator entries, $X^{\#}=[x_{ij}^*]$, $X^{\top}=[x_{ji}]$ is the usual transpose, and $X^{\dag}=(X^{\#})^{\top}$. The commutator of two operators $X$ and $Y$ is defined as $[X,Y]=XY-YX$.
  \item The identity matrix in $n$ dimensions will be denoted by $I_n$, and a $r \times s$ matrix of zeros will be denoted by $0_{r \times s}$. $\delta_{ij}$ denotes the Kronecker delta symbol, i.e. $I=[\delta_{ij}]$. We define $\mathbb{J}_{2k}=\left(\begin{smallmatrix} 0_{k \times k} & I_k \\ -I_k & 0_{k \times k} \end{smallmatrix}\right)$. Also, $\left(\begin{smallmatrix} X_1 \\ X_2 \\ \vdots \\ X_k \\ \end{smallmatrix}\right)$ is the vertical concatenation of the matrices $X_1,X_2,\ldots,X_k$, of equal column dimension, $(\,Y_1 \, Y_2 \, \ldots \, Y_k \,)$ is the horizontal concatenation of the matrices $Y_1, Y_2, \ldots, Y_k$ of equal row dimension, and $\diag(Z_1,Z_2,\ldots,Z_k)$ is the block-diagonal matrix formed by the square matrices $Z_1,Z_2,\ldots,Z_k$.
  \item For a $2r \times 2s$ matrix $X$, define its $\sharp$-\emph{adjoint} $X^{\sharp}$, by $X^{\sharp}= -\mathbb{J}_{2s}X^{\dag}\mathbb{J}_{2r}$. The $\sharp$-\emph{adjoint} satisfies properties similar to the usual adjoint, namely $(x_1 A + x_2 B)^{\sharp}=x_1^* A^{\sharp} + x_2^* B^{\sharp}$, $(AB)^{\sharp}=B^{\sharp}  A^{\sharp}$, and $(A^{\sharp})^{\sharp}=A$.
  \item A $2k \times 2k$ complex matrix $T$ is called \emph{symplectic}, if it satisfies $TT^{\sharp}=T^{\sharp}T=I_{2k}$. Hence, any symplectic matrix is invertible, and its inverse is its $\sharp$-adjoint. The set of these matrices forms a non-compact group known as the symplectic group.
\end{enumerate}

\subsection{Linear Quantum Stochastic Systems}
\label{Linear Quantum Stochastic Systems}

The material in this Subsection is fairly standard, and our presentation aims mostly at establishing notation and terminology. To this end, we follow the papers \cite{pet10,shapet12}. For the mathematical background necessary for a precise discussion of LQSSs, some standard references are \cite{par99,mey95,hudpar84}, while for a Physics perspective, see \cite{garzol00,garcol85}. The references \cite{nurjamdoh09,edwbel05,goujam09,gougohyan08,goujamnur10} contain a lot of relevant material, as well.

The systems we consider in this work are collections of quantum harmonic oscillators interacting among themselves, as well as with their environment. The $i$-th harmonic oscillator ($i=1,\ldots,n$) is described by its position and momentum variables, $q_i$ and $p_i$, respectively. These are self-adjoint operators satisfying the \emph{Canonical Commutation Relations} (CCRs) $[q_i,q_j]=0$, $[p_i,p_j]=0$, and $[q_i,p_j]=\imath\delta_{ij}$, for $i,j=1,\ldots,n$. As in classical mechanics, the states $q_{i}$ and $p_{i}$, $i=1,\ldots,n$, are called \emph{conjugate} states.  If we define the vectors of operators $q=(q_1,q_2,\dots,q_n)^{\top}$, $p=(p_1,p_2,\ldots,p_n)^{\top}$, and $x=\bigl(\begin{smallmatrix} q \\ p \end{smallmatrix} \bigr)$, the CCRs can be expressed as
\begin{eqnarray}
[x,x^{\top}] \doteq x x^{\top} -(x x^{\top})^{\top} =
\left(\begin{array}{cc}
\mathbf{0} & \imath I_n \\
-\imath I_n & \mathbf{0} \\
\end{array}\right)
= \imath \mathbb{J}_{2n}. \label{CCRs}
\end{eqnarray}

The environment is modelled as a collection of bosonic heat reservoirs. The $i$-th heat reservoir ($i=1,\ldots,m$) is described by bosonic \emph{field annihilation and creation operators} $\mathcal{A}_i(t)$ and $\mathcal{A}_i^*(t)$, respectively. The field operators are \emph{adapted quantum stochastic processes} with forward differentials $d\mathcal{A}_i(t)= \mathcal{A}_i(t+dt)-\mathcal{A}_i(t)$, and $d\mathcal{A}_i^*(t)= \mathcal{A}_i^*(t+dt)-\mathcal{A}_i^*(t)$. They satisfy the quantum It\^{o} products $d\mathcal{A}_i(t) d\mathcal{A}_j(t)=0$, $d\mathcal{A}_i^*(t) d\mathcal{A}_j^*(t)=0$, $d\mathcal{A}_i^*(t) d\mathcal{A}_j(t)=0$, and $d\mathcal{A}_i(t) d\mathcal{A}_j^*(t)=\delta_{ij} dt$. If we define the vector of field operators $\mathcal{A}(t)=(\mathcal{A}_1(t),\mathcal{A}_2(t),\dots,\mathcal{A}_m(t))^{\top}$, and the vector of self-adjoint field quadratures
\[ \mathcal{V}(t)=\frac{1}{\sqrt{2}}\left(\begin{array}{c}
\mathcal{A}(t)+\mathcal{A}(t)^{\#} \\ \imath(\mathcal{A}(t)-\mathcal{A}(t)^{\#}) \\
\end{array}\right), \]
the quantum It\^{o} products above can be expressed as
\setlength{\arraycolsep}{1pt}
\begin{eqnarray}
d\mathcal{V}(t) d\mathcal{V}(t)^{\top}=\frac{1}{2} \left(\!\!\begin{array}{cc}
I_m & \imath I_m \\
-\imath I_m & I_m \\
\end{array}\!\!\right) dt= \frac{1}{2}\,(I_{2m} + \imath \mathbb{J}_{2m}) dt. \label{Quantum Ito products}
\end{eqnarray}
\setlength{\arraycolsep}{3pt}

To describe the dynamics of the harmonic oscillators and the quantum fields, we introduce certain operators. We begin with the Hamiltonian operator $H=\frac{1}{2}x^{\top}Rx$, which specifies the dynamics of the harmonic oscillators in the absence of any environmental influence. $R$ is a $2n \times 2n$ real symmetric matrix referred to as the Hamiltonian matrix. Next, we have the coupling operator $L$ (vector of operators) that specifies the interaction of the harmonic oscillators with the quantum fields. $L$ depends linearly on the position and momentum operators of the oscillators, and can be expressed  as $L=L_q q + L_p p$. We construct the real coupling matrix $C^{2m \times 2n}$ from $L_q^{m \times n}$ and $L_p^{m \times n}$, as $C=\frac{1}{\sqrt{2}}\bigl(\begin{smallmatrix} L_q + L_q^{\#} & L_p + L_p^{\#} \\ -\imath(L_q - L_q^{\#}) & -\imath(L_p - L_p^{\#}) \end{smallmatrix}\bigr)$. Finally, we have the unitary scattering matrix $S^{m \times m}$, that describes the interactions between the quantum fields themselves.

In the \emph{Heisenberg picture} of quantum mechanics, the joint evolution of the harmonic oscillators and the quantum fields is described by the following system of \emph{Quantum Stochastic Differential Equations} (QSDEs):
\begin{eqnarray}
dx &=& (\mathbb{J} R -\frac{1}{2}C^{\sharp}C) x dt -C^{\sharp}\Sigma\, d\mathcal{V}, \nonumber \\
d\mathcal{V}_{out}&=& C x dt + \Sigma\, d\mathcal{V}, \label{General LQSS in real form}
\end{eqnarray}
where
\[ \Sigma = \frac{1}{2} \left(\begin{array}{cc}
S+S^{\#} & \imath (S-S^{\#}) \\
-\imath (S-S^{\#}) & S+S^{\#} \\
\end{array}\right), \]
is a $2m \times 2m$ real orthogonal symplectic matrix. The field quadrature operators $\mathcal{V}_{i \, out}(t)$ describe the outputs of the system. (\ref{General LQSS in real form}) is a description of the dynamics of the LQSS in the real quadrature operator representation, where the states, inputs, and outputs are all self-adjoint operators. We are going to use a version of (\ref{General LQSS in real form}) generalized in two ways: First, we replace the real orthogonal symplectic transformation $\Sigma$,  with a more general real symplectic transformation $\Sigma$, see e.g. \cite{goujamnur10} for a discussion in the creation-annihilation representation. Second, in the context of coherent quantum systems in particular, the output of a quantum system may be fed into other quantum system, so we substitute the more general input and output notations $\mathcal{U}$ and $\mathcal{Y}$, for $\mathcal{V}$ and $\mathcal{V}_{out}$, respectively. The resulting QSDEs are the following:
\begin{eqnarray}
dx &=& (\mathbb{J} R -\frac{1}{2}C^{\sharp}C) x dt -C^{\sharp}\Sigma\, d\mathcal{U}, \nonumber \\
d\mathcal{Y}&=& C x dt + \Sigma\, d\mathcal{U}, \label{General LQSS in real form 2}
\end{eqnarray}
The forward differentials $d\mathcal{U}$ and $d\mathcal{Y}$ of inputs and outputs, respectively (or, more precisely, of their quadratures), contain ``quantum noises'', as well as a ``signal part'' (linear combinations of variables of other systems). One can prove that, the structure of  (\ref{General LQSS in real form 2}) is preserved under linear transformations of the state $\bar{x}= T x$, if and only if $T$ is real symplectic (with $\bar{R}=T^{-\top} R T^{-1}$, and $\bar{C}=CT^{-1}=CT^{\#}$). From the point of view of quantum mechanics, $T$ must be real symplectic so that the transformed position and momentum operators are also self-adjoint and satisfy the same CCRs, as one can verify from (\ref{CCRs}). It is exactly this additional constraint on the allowed state transformations of LQSSs that complicates the construction of the Kalman decomposition for these systems.

\section{The Kalman Decomposition for Linear Quantum Stochastic Systems}
\label{The Kalman Decomposition for Linear Quantum Stochastic Systems}

System (\ref{General LQSS in real form 2}) has the standard form of a linear, time-invariant, system with $A=\mathbb{J}_{2n} R -\frac{1}{2}C^{\sharp}C$, $B=-C^{\sharp}\Sigma$, and $D=\Sigma$. However, as discussed in Subsection \ref{Linear Quantum Stochastic Systems}, only linear transformations of the state $\bar{x}= T x$, with $T$ real symplectic, preserve its structure, or, equivalently, preserve the self-adjointness and the CCRs of the states. In the following, we prove that there exists a real symplectic transformation of the state that puts (\ref{General LQSS in real form 2}) in a Kalman-like canonical form. Before we state and prove this result, we introduce the conventions used in this work regarding the uncontrollable and observable subspaces. Let
\begin{eqnarray}
\mathcal{C} &=& \left(\begin{array}{cccc}
B & AB & \cdots & A^{2n-1}B \\
\end{array}\right), \ \mathrm{and} \nonumber \\
\mathcal{O} &=& \left(\begin{array}{c}
C \\
CA \\
\vdots \\
CA^{2n-1} \\ \end{array}\right), \label{Controllability and Observability matrices}
\end{eqnarray}
be the controllability and observability matrices of the system (\ref{General LQSS in real form 2}). As usual, $\Ima \mathcal{C}$, and $\Ker \mathcal{O}$ define the controllable and unobservable subspaces. The uncontrollable and observable subspaces are defined as the \emph{orthogonal complements} of $\Ima \mathcal{C}$, and $\Ker \mathcal{O}$ in $\mathbb{R}^{2n}$, respectively. With this convention, we have the following theorem:
\begin{theorem}\label{Quantum Kalman Decomposition}
Given the LQSS (\ref{General LQSS in real form 2}), there exists a real symplectic transformation $V$ such that the following hold:
\begin{enumerate}
  \item The transformed states $\bigl(\begin{smallmatrix} \hat{q} \\ \hat{p} \end{smallmatrix}\bigr)=\hat{x} = Vx= V \bigl(\begin{smallmatrix} q \\ p \end{smallmatrix}\bigr)$, can be partitioned as follows:
\begin{eqnarray} \label{State partition}
\hat{q}=
\left(\begin{array}{l}
\hat{q}^{k \times 1}_a \\
\hat{q}^{l \times 1}_b \\
\hat{q}^{(n-k-l) \times 1}_c \\
\end{array}\right),
\hat{p}=
\left(\begin{array}{l}
\hat{p}^{k \times 1}_a \\
\hat{p}^{l \times 1}_b \\
\hat{p}^{(n-k-l) \times 1}_c \\
\end{array}\right),
\end{eqnarray}
  where
\begin{enumerate}
  \item The states $\hat{q}_a$ and $\hat{p}_a$ are both controllable and observable.
  \item The states $\hat{p}_b$ are controllable but unobservable.
  \item The states $\hat{q}_b$ are uncontrollable but observable.
  \item The states $\hat{q}_c$ and $\hat{p}_c$ are both uncontrollable and unobservable.
\end{enumerate}
  \item In the transformed states, (\ref{General LQSS in real form 2}) takes the form
\begin{eqnarray}
d\hat{x} &=& \hat{A}\hat{x}dt + \hat{B}d\mathcal{U}, \nonumber \\
d\mathcal{Y} &=& \hat{C}\hat{x}dt + D d\mathcal{U}, \label{Modified Kalman canonical form}
\end{eqnarray}
where
\small
\setlength{\arraycolsep}{2pt}
\begin{eqnarray*}
\hat{A}=\left(\begin{array}{cccccc}
A_{co,11} & A_{13,1} & 0 & A_{co,12} & 0 & 0 \\
0 & A_{\bar{c}o} & 0 & 0 & 0 & 0 \\
0 & A_{43,1} & A_{\bar{c}\bar{o},11} & 0 & 0 & A_{\bar{c}\bar{o},12} \\
A_{co,21} & A_{13,2} & 0 & A_{co,22} & 0 & 0 \\
A_{21,1} & A_{23} & A_{24,1} & A_{21,2} & A_{c\bar{o}} & A_{24,2} \\
0 & A_{43,2} & A_{\bar{c}\bar{o},21} & 0 & 0 & A_{\bar{c}\bar{o},22} \\
\end{array}\right),
\end{eqnarray*}
\normalsize
and
\small
\begin{eqnarray}
\hat{B} = \left(\!\begin{array}{c} B_{co,1} \\ 0 \\ 0 \\ B_{co,2} \\ B_{c\bar{o}} \\ 0 \\ \end{array}\!\right),
\hat{C} = \left(\begin{array}{cccccc}
C_{co,1} & C_{\bar{c}o} & 0 & C_{co,2} & 0 & 0 \end{array}\right) . \square \label{System matrices in modified Kalman canonical form}
\end{eqnarray}
\setlength{\arraycolsep}{3pt}
\normalsize
\end{enumerate}
\end{theorem}
To prove Theorem \ref{Quantum Kalman Decomposition}, we shall need the following lemmas:
\begin{lemma} \label{New Controllability and Observability matrices}
Let
\begin{eqnarray}
\tilde{\mathcal{C}} &=& \left(\begin{array}{cccc}
B & (\mathbb{J}R)B & \cdots & (\mathbb{J}R)^{2n-1}B \\
\end{array}\right), \ \mathrm{and} \nonumber \\
\tilde{\mathcal{O}} &=& \left(\begin{array}{c}
C \\
C(\mathbb{J}R) \\
\vdots \\
C(\mathbb{J}R)^{2n-1} \\ \end{array}\right). \label{New Controllability and Observability matrices definition}
\end{eqnarray}
Then, $\Ima \tilde{\mathcal{C}}=\Ima \mathcal{C}$, and $\Ker \tilde{\mathcal{O}}=\Ker \mathcal{O}$. $\square$
\end{lemma}
This follows from standard results of linear systems theory, since the system (\ref{General LQSS in real form 2}) can be constructed from a system with $(\tilde{A}, \tilde{B}, \tilde{C}, \tilde{D}) = (\mathbb{J}R, B, C, D)$, with state feedback with gain $\frac{1}{2}D^{-1}C$, or from a system with $(\tilde{A}, \tilde{B}, \tilde{C}, \tilde{D}) = (\mathbb{J}R, \frac{1}{2}B, C, D)$ with output injection with gain $-\frac{1}{2}C^{\sharp}$. Hence, in all of the constructions above, we may use $\tilde{\mathcal{C}}$ and $\tilde{\mathcal{O}}$ in place of $\mathcal{C}$ and $\mathcal{O}$. From now on, we shall refer to $\tilde{\mathcal{C}}$ and $\tilde{\mathcal{O}}$ simply as the controllability and observability matrices of the system (\ref{General LQSS in real form 2}). Next, we need another simple fact from linear systems theory:
\begin{lemma} \label{Transformation of Controllability and Observability matrices}
The controllability and observability matrices of a linear time-invariant control system $\mathcal{F}$, $\mathcal{C}_{\mathcal{F}}$ and $\mathcal{O}_{\mathcal{F}}$, respectively, transform as follows under a linear transformation of the state $x_{new}=V x$:
\begin{equation}\label{Transformations of Controllability and Observability matrices}
\mathcal{C}_{\mathcal{F},new}=V\mathcal{C}_{\mathcal{F}}, \ \ \mathcal{O}_{\mathcal{F},new}=\mathcal{O}_{\mathcal{F}}V^{-1}.\ \square
\end{equation}
\end{lemma}
The third result we shall make use of, is the following:
\begin{lemma} \label{Relation between Controllability and Observability matrices}
There exists a symplectic matrix $T_0$, such that $\tilde{\mathcal{O}}=T_0 \, \tilde{\mathcal{C}}^{\sharp}$, or, equivalently, $\tilde{\mathcal{C}}=\tilde{\mathcal{O}}^{\sharp}T_0$. $\square$
\end{lemma}
\textbf{Proof:} Let $X_1, X_2, \ldots, X_k$ be complex matrices of corresponding dimensions $2r \times 2s_1, \ldots, 2r \times 2s_k$. Then,
\begin{eqnarray*}
&&\left(\!\begin{array}{ccc} X_1 & \cdots & X_k \\ \end{array}\!\right)^{\sharp} = -\mathbb{J}_{2(s_1+ \ldots +s_k)}
\left(\!\begin{array}{ccc} X_1 & \cdots & X_k \\ \end{array}\!\right)^{\dag} \mathbb{J}_{2r} \\
&=&-\mathbb{J}_{2(s_1+ \ldots +s_k)}\left(\begin{array}{c}
X_1^{\dag}\mathbb{J}_{2r} \\
\vdots \\
X_k^{\dag}\mathbb{J}_{2r} \\
\end{array}\right) \\
&=& -\mathbb{J}_{2(s_1+ \ldots +s_k)}
\left(\!\!\begin{array}{ccc}
\mathbb{J}_{2s_1} &  & \mbox{\huge{0}} \\
 & \ddots &  \\
\mbox{\huge{0}} &  & \mathbb{J}_{2s_k} \\
\end{array}\!\!\right)
\left(\!\!\begin{array}{c}
-\mathbb{J}_{2s_1} X_1^{\dag} \mathbb{J}_{2r} \\
\vdots \\
-\mathbb{J}_{2s_1} X_k^{\dag} \mathbb{J}_{2r} \\
\end{array}\!\!\right) \\
&=& -\mathbb{J}_{2(s_1+ \ldots +s_k)}\, \diag(\mathbb{J}_{2s_1},\ldots,\mathbb{J}_{2s_k}) \left(\begin{array}{c}
X_1^{\sharp} \\
\vdots \\
X_k^{\sharp} \\
\end{array}\right).
\end{eqnarray*}
Applying the above result to $\tilde{\mathcal{C}}$, we have that
\begin{eqnarray*}
\tilde{\mathcal{C}}^{\sharp} &=& -\mathbb{J}_{4nm}\, \diag(\underbrace{\mathbb{J}_{2m},\ldots,\mathbb{J}_{2m}}_{2n})
\left(\!\!\begin{array}{c}
B^{\sharp} \\
((\mathbb{J}R)\, B)^{\sharp} \\
\vdots \\
((\mathbb{J}R)^{2n-1}B)^{\sharp} \\ \end{array}\!\!\right) \\
&=& -\mathbb{J}_{4nm}\, \diag(\mathbb{J}_{2m},\ldots,\mathbb{J}_{2m})
\left(\!\!\begin{array}{c}
B^{\sharp} \\
B^{\sharp} (\mathbb{J}R)^{\sharp} \\
\vdots \\
B^{\sharp} ((\mathbb{J}R)^{\sharp})^{2n-1} \\ \end{array}\!\!\right).
\end{eqnarray*}
However, $B^{\sharp}=(-C^{\sharp}D)^{\sharp}=-D^{\sharp}C=-D^{-1}C$, since $T^{\sharp}=T^{-1}$ for a symplectic $T$, and $(\mathbb{J} R)^{\sharp}= R^{\sharp} \mathbb{J}^{\sharp} = (-\mathbb{J} R^{\dag} \mathbb{J})\, (-\mathbb{J})=-\mathbb{J} R^{\dag}=-\mathbb{J} R$, due to the fact that $R$ is real symmetric. Putting everything together, we have that
\begin{eqnarray*}
\tilde{\mathcal{C}}^{\sharp} &=& -\mathbb{J}_{4nm}\, \diag(\mathbb{J}_{2m},\ldots,\mathbb{J}_{2m}) \, \diag(D^{-1},\ldots,D^{-1}) \\
&\times& \left(\begin{array}{c}
-C \\
-C (-\mathbb{J}R) \\
\vdots \\
-C (-\mathbb{J}R)^{2n-1} \\ \end{array}\right) = T_0^{-1} \tilde{\mathcal{O}},
\end{eqnarray*}
where
\begin{eqnarray*}
T_0^{-1} &=& \mathbb{J}_{4nm}\, \diag(\mathbb{J}_{2m},-\mathbb{J}_{2m},\ldots,\mathbb{J}_{2m},-\mathbb{J}_{2m}) \\
&\times& \diag(D^{-1},\ldots,D^{-1}).
\end{eqnarray*}
Since each of the matrices $\mathbb{J}_{4nm}$, $\diag(\mathbb{J}_{2m},-\mathbb{J}_{2m},\ldots,\mathbb{J}_{2m}$, $-\mathbb{J}_{2m})$, and  $\diag(D^{-1},\ldots,D^{-1})$ is real symplectic, the conclusion of the lemma follows with
\begin{eqnarray*}
T_0 &=& \diag(D,\ldots,D) \\
&\times& \diag(\mathbb{J}_{2m},-\mathbb{J}_{2m},\ldots,\mathbb{J}_{2m},-\mathbb{J}_{2m}) \, \mathbb{J}_{4nm}. \blacksquare
\end{eqnarray*}
The final result we need is the following ``one-sided symplectic'' SVD from \cite{xu03}:
\begin{lemma}\cite[Theorem 3]{xu03} \label{One-sided symplectic SVD thm}
For any matrix $F \in \mathbb{R}^{s \times 2r}$, there exist an orthogonal matrix $Q^{s \times s}$, and a real symplectic matrix $Z^{2r \times 2r}$, such that
\begin{equation}\label{One-sided symplectic SVD eqn1}
F=Q\, E \, Z^{-1},
\end{equation}
where
\setlength{\arraycolsep}{1pt}
\begin{eqnarray}\label{One-sided symplectic SVD eqn2}
&& E^{s \times 2r} = \nonumber \\
&& \begin{blockarray}{ccccccl}
k & l & r-k-l & k & l & r-k-l & \\
\begin{block}{(cccccc)l}
\Xi_k & 0 & 0 & 0 & 0 & 0 & k \\
0 & I_l & 0 & 0 & 0 & 0 & l \\
0 & 0 & 0 & \Xi_k & 0 & 0 & k \\
0 & 0 & 0 & 0 & 0 & 0 & l', \\
\end{block}
\end{blockarray}
\end{eqnarray}
\setlength{\arraycolsep}{3pt}
with $l'=s-2k-l$, and $\Xi_k=\diag(\xi_1,\ldots,\xi_k)>0$. $\square$
\end{lemma}
\textbf{Proof of Theorem \ref{Quantum Kalman Decomposition}:} We begin by applying Lemma \ref{One-sided symplectic SVD thm} to the observability matrix $\tilde{\mathcal{O}}^{4nm \times 2n}$ of system (\ref{General LQSS in real form 2}). Then, $\tilde{\mathcal{O}}=Q\, E \, Z^{-1}$ as above, with $s=4nm$ and $r=n$, while the integers $k$ and $l$ are determined by the lemma. Using Lemma \ref{Relation between Controllability and Observability matrices}, we have that $\tilde{\mathcal{C}} = \tilde{\mathcal{O}}^{\sharp}T_0 = (Q\, E \, Z^{-1})^{\sharp}T_0 = (Z^{-1})^{\sharp} E^{\sharp} Q^{\sharp}T_0 = Z\, E^{\sharp} Q^{\sharp}T_0$. Now, we perform the state transformation $\bigl(\begin{smallmatrix} \hat{q} \\ \hat{p} \end{smallmatrix}\bigr) = Z^{-1} \bigl(\begin{smallmatrix} q \\ p \end{smallmatrix}\bigr)$. Since $Z$ and $Z^{-1}$ are real symplectic, the transformed system is also of the form (\ref{General LQSS in real form 2}). According to Lemma \ref{Transformation of Controllability and Observability matrices}, the controllability and observability matrices of the transformed system are given by
\begin{eqnarray}
\hat{\tilde{\mathcal{C}}} &=& Z^{-1} \tilde{\mathcal{C}}=Z^{-1} Z\, E^{\sharp} Q^{\sharp}T_0 = E^{\sharp} Q^{\sharp}T_0, \label{Transformed Controllability matrix} \\
\hat{\tilde{\mathcal{O}}} &=& \tilde{\mathcal{O}}\, (Z^{-1})^{-1} = Q\, E \, Z^{-1} Z = Q\, E. \label{Transformed Observability matrix}
\end{eqnarray}
Since $Q$ is of full rank, (\ref{Transformed Observability matrix}) implies that $\Ker \hat{\tilde{\mathcal{O}}}= \Ker E$. Let $e_i$ denote the $i$-th vector of the standard basis of $\mathbb{R}^{2n}$. Then, we conclude that
\begin{eqnarray*}
&& \Ker \hat{\tilde{\mathcal{O}}}= \Ker E \\
&=& \Span\{e_{k+l+1},\ldots,e_n,e_{n+k+1},\ldots,e_{2n}\}.
\end{eqnarray*}
From (\ref{Transformed Controllability matrix}), we have that
\begin{eqnarray*}
&& \Ima \hat{\tilde{\mathcal{C}}}= \Ima E^{\sharp} Q^{\sharp}T_0 = \Ima E^{\sharp}= \Ima (-\mathbb{J}E^{\top}\mathbb{J})= \Ima \mathbb{J}E^{\top} \\[.1em]
&=& \Ima \left(\begin{array}{cccc}
0 & 0 & \Xi_k & 0 \\
0 & 0 & 0 & 0 \\
0 & 0 & 0 & 0 \\
-\Xi_k & 0 & 0 & 0 \\
0 & -I_l & 0 & 0 \\
0 & 0 & 0 & 0 \\
\end{array}\right)= \Span \{e_1,\ldots,e_k, \\[.1em]
&& e_{n+1},\ldots,e_{n+k},e_{n+k+1},\ldots, e_{n+k+l}\}.
\end{eqnarray*}
The fact that $Q$ and $T_0$ are of full rank was used in the above derivation. If we partition the states as in (\ref{State partition}),
\begin{eqnarray*}
\hat{q}=
\left(\begin{array}{l}
\hat{q}^{k \times 1}_a \\
\hat{q}^{l \times 1}_b \\
\hat{q}^{(n-k-l) \times 1}_c \\
\end{array}\right), \ \mathrm{and}\
\hat{p}=
\left(\begin{array}{l}
\hat{p}^{k \times 1}_a \\
\hat{p}^{l \times 1}_b \\
\hat{p}^{(n-k-l) \times 1}_c \\
\end{array}\right),
\end{eqnarray*}
the calculations of the controllable and unobservable subspaces above, along with our convention for the uncontrollable and observable subspaces, lead to the following picture:
\begin{enumerate}
  \item The states $\hat{q}_a$, $\hat{p}_a$, and $\hat{p}_b$ are controllable, and the  states $\hat{q}_b$, $\hat{q}_c$, and $\hat{p}_c$ are uncontrollable.
  \item The states $\hat{q}_c$, $\hat{p}_b$, and $\hat{p}_c$ are unobservable, and the states $\hat{q}_a$, $\hat{q}_b$, and $\hat{p}_a$ are observable.
\end{enumerate}
Combining the above controllability and observability results, we end up with the classification of states announced in the statement of the theorem.

Hence, the state transformation $\bigl(\begin{smallmatrix} \hat{q} \\ \hat{p} \end{smallmatrix}\bigr) = V \bigl(\begin{smallmatrix} q \\ p \end{smallmatrix}\bigr)$, with $V=Z^{-1}$, essentially puts the system in the Kalman canonical form. The qualification has to do with the fact that, the usual grouping of states in the Kalman canonical form, $(x_{co}, x_{c\bar{o}}, x_{\bar{c}o}, x_{\bar{c}\bar{o}})$, is incompatible with the grouping of the states of (\ref{General LQSS in real form 2}) in conjugate pairs of position and momentum coordinates, $(\hat{q},\hat{p})$, that is necessary for the structure of (\ref{General LQSS in real form 2}) to be preserved. The resolution of this issue is, to modify the usual Kalman canonical form. To do this, we start from the usual Kalman canonical form \cite{zhodoyglo96,kim97}
\begin{eqnarray*}
d \left(\begin{array}{c}
x_{co} \\
x_{c\bar{o}} \\
x_{\bar{c}o} \\
x_{\bar{c}\bar{o}} \\
\end{array}\right) &=&
\left(\begin{array}{cccc}
A_{co} & 0 & A_{13} & 0 \\
A_{21} & A_{c\bar{o}} & A_{23} & A_{24} \\
0 & 0 & A_{\bar{c}o} & 0 \\
0 & 0 & A_{43} & A_{\bar{c}\bar{o}} \\
\end{array}\right)
\left(\begin{array}{c}
x_{co} \\
x_{c\bar{o}} \\
x_{\bar{c}o} \\
x_{\bar{c}\bar{o}} \\
\end{array}\right) dt  \\
&+&\left(\begin{array}{c}
B_{co} \\
B_{c\bar{o}} \\
0 \\
0 \\
\end{array}\right) d\mathcal{U},
\end{eqnarray*}
\begin{eqnarray}
d\mathcal{Y} =
\left(\begin{array}{cccc}
C_{co} & 0 & C_{\bar{c}o} & 0 \\
\end{array}\right) \left(\begin{array}{c}
x_{co} \\
x_{c\bar{o}} \\
x_{\bar{c}o} \\
x_{\bar{c}\bar{o}} \\
\end{array}\right) dt + D\, d\mathcal{U}, \label{System in Kalman canonical form}
\end{eqnarray}
and let $x_{co}= \bigl(\begin{smallmatrix} \hat{q}_a \\ \hat{p}_a \end{smallmatrix}\bigr)$, $x_{c\bar{o}}=\hat{p}_b$, $x_{\bar{c}o}=\hat{q}_b$, $x_{\bar{c}\bar{o}}=\bigl(\begin{smallmatrix} \hat{q}_c \\ \hat{p}_c \end{smallmatrix}\bigr)$. Also, partition $A_{co}=\bigl(\begin{smallmatrix}
A_{co,11} & A_{co,12} \\ A_{co,21}& A_{co,22} \end{smallmatrix}\bigr)$, $A_{13}=\bigl(\begin{smallmatrix} A_{13,1} \\ A_{13,2} \end{smallmatrix}\bigr)$, $A_{21}=\bigl(\begin{smallmatrix} A_{21,1} & A_{21,2} \end{smallmatrix}\bigr)$, $A_{24}=\bigl(\begin{smallmatrix} A_{24,1} & A_{24,2} \end{smallmatrix}\bigr)$, $A_{43}=\bigl(\begin{smallmatrix} A_{43,1} \\ A_{43,2} \end{smallmatrix}\bigr)$, $A_{\bar{c}\bar{o}}=\bigl(\begin{smallmatrix} A_{\bar{c}\bar{o},11} & A_{\bar{c}\bar{o},12} \\ A_{\bar{c}\bar{o},21}& A_{\bar{c}\bar{o},22} \end{smallmatrix}\bigr)$, $B_{co}=\bigl(\begin{smallmatrix} B_{co,1} \\ B_{co,2} \end{smallmatrix}\bigr)$, and $C_{co}=\bigl(\begin{smallmatrix} C_{co,1} & C_{co,2} \end{smallmatrix}\bigr)$, accordingly. Then, by reshuffling the Kalman canonical form, we end up with (\ref{Modified Kalman canonical form}), where $\hat{A}$, $\hat{B}$, and $\hat{C}$ are given by (\ref{System matrices in modified Kalman canonical form}). $\blacksquare$

Though Theorem \ref{Quantum Kalman Decomposition} constructs one particular Kalman decomposition of the LQSS (or, equivalently, one particular Kalman-like canonical form (\ref{Modified Kalman canonical form})\,), it is easy to generate many more by use of the following corollary:
\begin{coro} \label{Possible Kalman Transformations}
Let $E \in \mathbb{R}^{4nm \times 2n}$ be the reduced form of the observability matrix $\tilde{\mathcal{O}}\in \mathbb{R}^{4nm \times 2n}$ of system (\ref{General LQSS in real form 2}), according to Lemma \ref{One-sided symplectic SVD thm}, see equation (\ref{One-sided symplectic SVD eqn2}). Also, let $X \in \mathbb{R}^{4nm \times 4nm}$ be invertible, and $Y \in \mathbb{R}^{2n \times 2n}$ symplectic, such that,
\begin{eqnarray}
&& X\,E\,Y \nonumber \\
&=&\begin{blockarray}{ccccccl}
k & l & n-k-l & k & l & n-k-l & \\
\begin{block}{(cccccc)l}
\Xi_k^{'} & 0 & 0 & 0 & 0 & 0 & k \\
0 & \Xi_l^{'} & 0 & 0 & 0 & 0 & l \\
0 & 0 & 0 & \Xi_k^{''} & 0 & 0 & k \\
0 & 0 & 0 & 0 & 0 & 0 & l', \\
\end{block}
\end{blockarray}
\end{eqnarray}
with $l'=4nm-2k-l$, and every element of the diagonal matrices $\Xi_k^{'} \in \mathbb{R}^{k \times k}$, $\Xi_l^{'} \in \mathbb{R}^{l \times l}$, and $\Xi_k^{''} \in \mathbb{R}^{k \times k}$, is non-zero. If $V$ is the symplectic transformation to the Kalman-like canonical form in Theorem \ref{Quantum Kalman Decomposition}, then the theorem holds for $V'=Y^{-1}V$, as well. $\square$
\end{coro}
\textbf{Proof:} We have that
\[\tilde{\mathcal{O}} = Q\, E \, Z^{-1} = (Q\, X^{-1})\, (X\,E\,Y)\,(Y^{-1}Z^{-1}).\]
In the proof of Theorem \ref{Quantum Kalman Decomposition}, the fact that $Q$ is unitary was used just to guarantee that it is of full rank. Also, the exact values of the elements of the non-zero diagonal blocks of $E$ were unimportant. It is straightforward to see that, the proof of the theorem follows through using the decomposition above, instead of (\ref{One-sided symplectic SVD eqn2}). The conclusion of the corollary follows. $\blacksquare$

\section{An Example}
\label{An Example}

Consider the following 3-mode, 1 input/output LQSS with Hamiltonian
\[H=\frac{\omega}{2}(q_3^2 + p_3^2) +\lambda q_1 q_3 + \lambda q_2 q_3,   \]
and coupling operator
\[L=\frac{\gamma}{\sqrt{2}}(q_3 + \imath p_3) . \]
This LQSS models the linearized dynamics of an optomechanical system where the resonant modes of two optical cavities, with states $(q_1,p_1)$ and $(q_2,p_2)$, respectively, interact with a mechanical mode with states $(q_3,p_3)$, of frequency $\omega$. We assume that the cavities are lossless, and that their interaction strengths with the mechanical oscillator are equal. The only source of damping in the system is mechanical. The system QSDEs (\ref{General LQSS in real form 2}), take the following form:
\begin{eqnarray*}
dq_1 &=& 0, \\
dq_2 &=& 0, \\
dq_3 &=& \big(-\frac{\ \gamma^2}{2}q_3 + \omega p_3\big) \,dt -\gamma d\mathcal{U}_1, \\
dp_1 &=& -\lambda q_3 dt, \\
dp_2 &=& -\lambda q_3 dt, \\
dp_3 &=& -\big(\lambda q_1 + \lambda q_2 + \omega q_3 + \frac{\ \gamma^2}{2}p_3 \big) \,dt -\gamma d\mathcal{U}_2, \\
d\mathcal{Y}_1 &=& \gamma q_3 \,dt + d\mathcal{U}_1, \\
d\mathcal{Y}_2 &=& \gamma p_3 \,dt + d\mathcal{U}_2.
\end{eqnarray*}
Recall that $\mathcal{U}_1$ and $\mathcal{U}_2$ are the two real quadratures of a single input, and similarly for the outputs.

Applying the ``one-sided symplectic'' SVD of \cite{xu03} to the observability matrix of the above LQSS, we obtain the symplectic transformation $V$ that puts the system in the Kalman-like canonical form (\ref{Modified Kalman canonical form}):
\begin{eqnarray*}
V &=& \left(\begin{array}{cccccc}
0 & 0 & 0 & 0 & 0 & 1 \\
-1 & -1 & 0 & 0 & 0 & 0 \\
\frac{1}{\sqrt{2}} & -\frac{1}{\sqrt{2}} & 0 & 0 & 0 & 0 \\
-\lambda a & -\lambda a & -1 & 0 & 0 & 0 \\
0 & 0 & 0 & -1/2 & -1/2 & \lambda a \\
0 & 0 & 0 & \frac{1}{\sqrt{2}} & -\frac{1}{\sqrt{2}} & 0
\end{array}\right),
\end{eqnarray*}
where
\[a=\omega \frac{\omega^8 + \omega^6 + \omega^4 + \omega^2 + 1}{\omega^{10} + \omega^8 + \omega^6 + \omega^4 + \omega^2 + 1} . \]
The new states of the system are given by
\begin{eqnarray*}
\left(\begin{array}{c}
\hat{q}_1 \\
\hat{q}_2 \\
\hat{q}_3 \\
\hat{p}_1 \\
\hat{p}_2 \\
\hat{p}_3 \\
\end{array}\right) =
\left(\begin{array}{c}
p_3 \\
-(q_1 + q_2) \\
\frac{1}{\sqrt{2}}(q_1 - q_2) \\
- q_3 -\lambda a(q_1 + q_2) \\
\lambda a \, p_3 - \frac{1}{2}(p_1 + p_2) \\
\frac{1}{\sqrt{2}}(p_1 - p_2)
\end{array}\right).
\end{eqnarray*}
$\hat{q}_1$ and $\hat{p}_1$ are the $co$ states, $\hat{q}_2$ and $\hat{p}_2$ are the $\bar{c}o$ and $c\bar{o}$ states, respectively, and $\hat{q}_3$ and $\hat{p}_3$ are the $\bar{c}\bar{o}$ states. This is confirmed by the system QSDEs in the transformed states, which take the following form:
\begin{eqnarray*}
d\hat{q}_1 &=& \big(-\frac{\ \gamma^2}{2}\hat{q}_1 + \lambda b\, \hat{q}_2 + \omega \hat{p}_1\big)\, dt -\gamma d\mathcal{U}_2, \\
d\hat{q}_2 &=& 0, \\
d\hat{q}_3 &=& 0, \\
d\hat{p}_1 &=& \big(-\omega \hat{q}_1+ \lambda a\frac{\ \gamma^2}{2}\hat{q}_2 -\frac{\ \gamma^2}{2}\hat{p}_1 \big)\, dt + \gamma d\mathcal{U}_1, \\
d\hat{p}_2 &=& \big(-\lambda a\frac{\ \gamma^2}{2}\hat{q}_1 + \lambda^2 a (b+1)\, \hat{q}_2 -\lambda b\, \hat{p}_1 \big)\, dt \\
&-& \gamma \lambda a\, d\mathcal{U}_2, \\
d\hat{p}_3 &=& 0, \\
d\mathcal{Y}_1 &=& \gamma(\lambda a \, \hat{q}_2 -\hat{p}_1)\, dt + d\mathcal{U}_1, \\
d\mathcal{Y}_2 &=& \gamma \hat{q}_1 \,dt + d\mathcal{U}_2,
\end{eqnarray*}
where $b=1/(\omega^{10} + \omega^8 + \omega^6 + \omega^4 + \omega^2 + 1)$. We can use Corollary \ref{Possible Kalman Transformations}, to produce a simpler Kalman decomposition of the system. Indeed, with
\[ Y=\left(\begin{array}{cccccc}
0&0&0&1&0&0\\
0&-\sqrt{2}&0&0&0&0\\
0&0&1&0&0&0\\
-1&-\sqrt{2}\lambda a&0&0&0&0\\
0&0&0&\lambda a&-\frac{1}{\sqrt{2}}&0\\
0&0&0&0&0&1\\
\end{array}\right), \]
and
\[ X=\diag(\left(\begin{array}{ccc}
0 & -\lambda \gamma a/\sqrt{b} & 1 \\
0 & 1 & 0 \\
1 & 0 & 0 \\
\end{array}\right),I_9), \]
we obtain the following orthogonal symplectic transformation $V'=Y^{-1}V$, that puts the system in the Kalman-like canonical form (\ref{Modified Kalman canonical form}):
\begin{eqnarray*}
V' &=& \left(\begin{array}{cccccc}
0 & 0 & 1 & 0 & 0 & 0 \\
\frac{1}{\sqrt{2}} & \frac{1}{\sqrt{2}} & 0 & 0 & 0 & 0 \\
\frac{1}{\sqrt{2}} & -\frac{1}{\sqrt{2}} & 0 & 0 & 0 & 0 \\
0 & 0 & 0 & 0 & 0 & 1 \\
0 & 0 & 0 & \frac{1}{\sqrt{2}} &\frac{1}{\sqrt{2}} & 0 \\
0 & 0 & 0 & \frac{1}{\sqrt{2}} & -\frac{1}{\sqrt{2}} & 0
\end{array}\right).
\end{eqnarray*}
The new states of the system are given by
\begin{eqnarray*}
\left(\begin{array}{c}
\hat{q}_1 \\
\hat{q}_2 \\
\hat{q}_3 \\
\hat{p}_1 \\
\hat{p}_2 \\
\hat{p}_3 \\
\end{array}\right) =
\left(\begin{array}{c}
q_3 \\
\frac{1}{\sqrt{2}}(q_1 + q_2) \\
\frac{1}{\sqrt{2}}(q_1 - q_2) \\
p_3 \\
\frac{1}{2}(p_1 + p_2) \\
\frac{1}{\sqrt{2}}(p_1 - p_2)
\end{array}\right).
\end{eqnarray*}
Again, $\hat{q}_1$ and $\hat{p}_1$ are the $co$ states, $\hat{q}_2$ and $\hat{p}_2$ are the $\bar{c}o$ and $c\bar{o}$ states, respectively, and $\hat{q}_3$ and $\hat{p}_3$ are the $\bar{c}\bar{o}$ states. This is confirmed by the system QSDEs in the transformed states, which take the following form:
\begin{eqnarray*}
d\hat{q}_1 &=& \big(-\frac{\ \gamma^2}{2}\hat{q}_1 + \omega \hat{p}_1\big)\, dt -\gamma d\mathcal{U}_1, \\
d\hat{q}_2 &=& 0, \\
d\hat{q}_3 &=& 0, \\
d\hat{p}_1 &=& \big(-\omega \hat{q}_1 - \sqrt{2}\lambda \hat{q}_2 -\frac{\ \gamma^2}{2}\hat{p}_1 \big)\, dt - \gamma d\mathcal{U}_2, \\
d\hat{p}_2 &=& - \sqrt{2}\lambda\hat{q}_1 \, dt, \\
d\hat{p}_3 &=& 0, \\
d\mathcal{Y}_1 &=& \gamma \hat{q}_1 \, dt + d\mathcal{U}_1, \\
d\mathcal{Y}_2 &=& \gamma \hat{p}_1 \, dt + d\mathcal{U}_2.
\end{eqnarray*}

\bibliographystyle{ieeetr}
\bibliography{C:/Users/Symeon/Documents/AAA/Work/Latex/MyBibliographies/Linear_Quantum_Stochastic_Systems,C:/Users/Symeon/Documents/AAA/Work/Latex/MyBibliographies/Books,C:/Users/Symeon/Documents/AAA/Work/Latex/MyBibliographies/My_papers,C:/Users/Symeon/Documents/AAA/Work/Latex/MyBibliographies/Miscellaneous_papers}

\begin{thebibliography}{10}

\bibitem{zhagripet16}
G.~Zhang, S.~Grivopoulos, I.~R. Petersen, and J.~E. Gough, ``{The Kalman
  decomposition for linear quantum systems},'' 2016.
\newblock Submitted to the IEEE Transactions on Automatic Control. Preprint
  available online at http://lanl.arxiv.org/abs/1606.05719.

\bibitem{xu03}
H.~Xu, ``An {SVD}-like matrix decomposition and its applications,'' {\em Linear
  Algebra and its Applications}, vol.~368, pp.~1 -- 24, 2003.

\bibitem{garzol00}
C.~Gardiner and P.~Zoller, {\em {Quantum Noise}}.
\newblock Springer-Verlag, Berlin, second~ed., 2000.

\bibitem{walmil08}
D.~Walls and G.~Milburn, {\em {Quantum Optics}}.
\newblock Springer-Verlag, 2nd~ed., 2008.

\bibitem{wismil10}
H.~Wiseman and G.~Milburn, {\em {Quantum Measurement and Control}}.
\newblock Cambridge University Press, 2010.

\bibitem{matjirper11}
A.~Matyas, C.~Jirauschek, F.~Peretti, P.~Lugli, , and G.~Csaba, ``Linear
  circuit models for on-chip quantum electrodynamics,'' {\em IEEE Transactions
  on Microwave Theory and Techniques}, vol.~59, pp.~65--71, 2011.

\bibitem{kerandku13}
J.~Kerckhoff, R.~W. Andrews, H.~S. Ku, W.~F. Kindel, K.~Cicak, R.~W. Simmonds,
  , and K.~W. Lehnert, ``Tunable coupling to a mechanical oscillator circuit
  using a coherent feedback network,'' {\em Physical Review X}, vol.~3,
  p.~021013, 2013.

\bibitem{tsacav10}
M.~Tsang and C.~M. Caves, ``Coherent quantum-noise cancellation for
  optomechanical sensors,'' {\em Physical Review Letters}, vol.~105, p.~123601,
  2010.

\bibitem{masheipir11}
F.~Massel, T.~T. Heikkila, J.~M. Pirkkalainen, S.~U. Cho, H.~Saloniemi, P.~J.
  Hakonen, , and M.~A. Sillanpaa, ``Microwave amplification with nanomechanical
  resonators,'' {\em Nature}, vol.~480, pp.~351--354, 2011.

\bibitem{hammab12}
R.~Hamerly and H.~Mabuchi, ``{Advantages of coherent feedback for cooling
  quantum oscillators},'' {\em Physical Review Letters}, vol.~109, p.~173602,
  2012.

\bibitem{donfiokuz12}
C.~Dong, V.~Fiore, M.~C. Kuzyk, , and H.~Wang, ``Optomechanical dark mode,''
  {\em Science}, vol.~338, no.~6114, pp.~1609--1613, 2012.

\bibitem{par99}
K.~Parthasarathy, {\em {An Introduction to Quantum Stochastic Calculus}}.
\newblock Birkhauser, 1999.

\bibitem{mey95}
P.~Meyer, {\em {Quantum Probability for Probabilists}}.
\newblock Springer, second~ed., 1995.

\bibitem{hudpar84}
R.~L. Hudson and K.~R. Parthasarathy, ``Quantum {I}t\^{o}'s formula and
  stochastic evolutions,'' {\em Communications in Mathematical Physics},
  vol.~93, pp.~301--323, 1984.

\bibitem{yankim03a}
M.~Yanagisawa and H.~Kimura, ``{Transfer function approach to quantum
  control-part I: dynamics of quantum feedback systems},'' {\em IEEE
  Transactions on Automatic Control}, vol.~48, no.~12, pp.~2107--2120, 2003.

\bibitem{yankim03b}
M.~Yanagisawa and H.~Kimura, ``{Transfer function approach to quantum
  control-part II: control concepts and applications},'' {\em IEEE Transactions
  on Automatic Control}, vol.~48, no.~12, pp.~2121--2132, 2003.

\bibitem{jamnurpet08}
M.~James, H.~I. Nurdin, and I.~Petersen, ``{$H^{\infty}$} control of linear
  quantum stochastic systems,'' {\em IEEE Transactions on Automatic Control},
  vol.~53, pp.~1787--1803, Sept 2008.

\bibitem{nurjampet09}
H.~I. Nurdin, M.~R. James, and I.~R. Petersen, ``Coherent quantum {LQG}
  control,'' {\em Automatica}, vol.~45, no.~8, pp.~1837 -- 1846, 2009.

\bibitem{maapet11b}
A.~I. Maalouf and I.~R. Petersen, ``{Coherent $H^{\infty}$ control for a class
  of annihilation operator linear quantum systems},'' {\em IEEE Transactions on
  Automatic Control}, vol.~56, no.~2, pp.~309--319, 2011.

\bibitem{zhajam12}
G.~Zhang and M.~James, ``Quantum feedback networks and control: a brief
  survey,'' {\em Chinese Science Bulletin}, vol.~57, no.~18, pp.~2200--2214,
  2012.

\bibitem{mab08}
H.~Mabuchi, ``{Coherent-feedback quantum control with a dynamic compensator},''
  {\em Physical Review A}, vol.~78, p.~032323, 2008.

\bibitem{critezsoh13}
O.~Crisafulli, N.~Tezak, D.~B.~S. Soh, M.~A. Armen, and H.~Mabuchi, ``{Squeezed
  light in an optical parametric oscillator network with coherent feedback
  quantum control},'' {\em Optics Express}, vol.~21, no.~15, pp.~3761--3774,
  2013.

\bibitem{pet10}
I.~R. Petersen, ``Quantum linear systems theory,'' in {\em Proceedings of the
  19th International Symposium on Mathematical Theory of Networks and Systems},
  (Budapest, Hungary), July 2010.

\bibitem{gouzha15}
J.~E. Gough and G.~Zhang, ``On realization theory of quantum linear systems,''
  {\em Automatica}, vol.~59, pp.~139--151, 2015.

\bibitem{gutyam16}
M.~Guta and N.~Yamamoto, ``System identification for passive linear quantum
  systems,'' {\em IEEE Transactions on Automatic Control}, vol.~61, no.~4,
  pp.~921--936, 2016.

\bibitem{shapet12}
A.~A.~J.~Shaiju and I.~R. Petersen, ``A frequency domain condition for the
  physical realizability of linear quantum systems,'' {\em IEEE Transactions on
  Automatic Control}, vol.~57, pp.~2033--2044, August 2012.

\bibitem{garcol85}
C.~Gardiner and M.~Collett, ``Input and output in damped quantum systems:
  {Q}uantum stochastic differential equations and the master equation,'' {\em
  Physical Review A}, vol.~31, no.~6, pp.~3761--3774, 1985.

\bibitem{nurjamdoh09}
H.~I. Nurdin, M.~R. James, and A.~C. Doherty, ``Network synthesis of linear
  dynamical quantum stochastic systems,'' {\em SIAM Journal on Control and
  Optimization}, vol.~48, no.~4, pp.~2686--2718, 2009.

\bibitem{edwbel05}
S.~C. Edwards and V.~P. Belavkin, ``Optimal quantum filtering and quantum
  feedback control,'' {\em arXiv:quant-ph/0506018}, August 2005.
\newblock Preprint.

\bibitem{goujam09}
J.~Gough and M.~James, ``The series product and its application to quantum
  feedforward and feedback networks,'' {\em IEEE Transactions on Automatic
  Control}, vol.~54, pp.~2530--2544, Nov 2009.

\bibitem{gougohyan08}
J.~E. Gough, R.~Gohm, and M.~Yanagisawa, ``Linear quantum feedback networks,''
  {\em Physical Review A}, vol.~78, p.~062104, Dec 2008.

\bibitem{goujamnur10}
J.~E. Gough, M.~R. James, and H.~I. Nurdin, ``Squeezing components in linear
  quantum feedback networks,'' {\em Physical Review A}, vol.~81, p.~023804, Feb
  2010.

\bibitem{zhodoyglo96}
K.~Zhou, J.~Doyle, and K.~Glover, {\em {Robust and Optimal Control}}.
\newblock Prentice Hall, 1996.

\bibitem{kim97}
H.~Kimura, {\em {Chain Scattering Approach to $H_{\infty}$-Control}}.
\newblock Birkh\"{a}user, 1997.

\end{thebibliography}
\end{document}